\documentstyle[aps,twocolumn]{revtex}
\begin{document}
\flushbottom
\draft
\title{{\bf Two-atom dark states in electromagnetic cavities}}
\author{G.\ J.\ Yang,$^1$ O.\ Zobay, and P.\ Meystre}
\address{
Optical Sciences Center, University of Arizona, Tucson, Arizona
85721\\
{\rm \  }
\\ \medskip}\author{\small\parbox{14.2cm}{\small \hspace*{3mm}
The center-of-mass motion of two two-level atoms coupled to a single
damped mode of an electromagnetic resonator is investigated. For the
case of one atom being initially excited and the cavity mode in the
vacuum state it is shown that the atomic time evolution is dominated by
the appearance of dark states. These states, in which the initial
excitation is stored in the internal atomic degrees of freedom and
the atoms become quantum mechanically entangled, are almost
immune against photon loss from the cavity. Various properties of the
dark states within and beyond the Raman-Nath approximation of atom optics
are worked out.
\\[3pt]PACS numbers: 42.50.Ct, 42.50.Fx, 42.50.Vk}}\address{}
\maketitle
\maketitle
\narrowtext

\section{Introduction}
\footnotetext[1]{Permanent address:
Department of Physics, Beijing Normal University,
Beijing 100875, China.}
Recent advances in cavity quantum electrodynamics have
significantly expanded our understanding of the interaction between
matter and the quantized electromagnetic field \cite{Mey92,Har92}.
A central topic in these studies is the theoretical and
experimental investigation of situations in which
a single atom interacts with a small number of modes of the radiation field
in high-$Q$ optical or microwave resonators. In such a setting, the
dynamical behavior of the atom is evidently very different from the
free-space situation and one can observe phenomena such as inhibited and
enhanced spontaneous emission \cite{Pur46,Kle81} or Rabi oscillations between
two electromagnetically coupled states \cite{Bru96}. A natural extension of 
these studies concerns the modification of the interaction between two atoms
in a cavity environment. As the interatomic interaction is ultimately
mediated by the electromagnetic field, one can expect drastic effects
also in this case. The interest in this problem has recently grown,
stimulated in part by the remarkable experiments of Refs.\ \cite{Eic93}
and \cite{DeV96}. For example, several recent articles have examined
the mutual coherence of the two atomic dipoles under various
circumstances \cite{Koc95,Mey97,Rud98,Yeo98}.

In a further study the modification of the near-resonant dipole-dipole
interaction between two atoms confined to a cavity was investigated in
detail \cite{Gol97}. As a main result it was shown that the familiar concept
of the dipole-dipole potential ceases to be meaningful under certain
circumstances. The purpose of the present paper is to continue and
extend this work, the emphasis now being put on the investigation
of the actual dynamical behavior of the atoms. In particular, we examine
the atomic center-of-mass motion under the influence of their interaction
with the cavity field. In order to work out basic aspects of the problem we
concentrate here on the model of a short and closed optical resonator
in which the atoms interact exclusively with a single damped standing-wave
mode of the electromagnetic radiation field. An initially excited atom will
then spontaneously emit a photon into the cavity mode and subsequently
reabsorb it. Consequently, it experiences a random walk in momentum space,
i.e.\ heating. Due to photon exchange the atom can also interact with and
excite its partner in the cavity. These processes will cease, of course, as
soon as the photon escapes the resonator due to cavity losses.

The analysis of this problem shows that, contrary to what one might
expect intuitively, the presence of the second atom does not simply lead
to some quantitative modifications in the heating and decay process of
the first. Rather, it causes qualitative changes in the dynamical behavior
of the system. In particular, one observes a tendency of the system to
settle into so-called ``dark" or ``quasi-dark" states. These dark states
consist of superpositions of states in which the initial excitation is stored
in either atom 1 or atom 2, i.e.,  entangled states of the atoms-cavity
system. Due to destructive quantum interferences
these superpositions are completely --- or to a large degree --- dynamically
decoupled from the states in which the photon is present in the cavity.
Thus they are immune --- or almost immune --- to photon decay. Atoms in
these dark states can be thought of as a new kind of ``molecule'' largely
delocalized and bound by the cavity electromagnetic field.
The focus of the present article lies on an analysis of these dark
states, which can be viewed as a generalization of the antisymmetric
Dicke state of the theory of super- and subradiance \cite{Dic54}. To our
knowledge, the persistence of the entangled two-atom dark states under
the influence of the atomic center-of-mass motion has not been previously
discussed in the literature. 

Section II introduces our model and establishes the notation. In order to 
motivate the subsequent analysis, Sec.~III discusses some numerical examples
that illustrate the role of the dark states and demonstrate their
long-livedness, even in the case of only approximate darkness. Section
IV gives a detailed analytical discussion of the dark states. We first 
consider the dynamics of the atomic system in the Raman-Nath approximation 
(RNA), where the atoms are treated as infinitely massive. This allows for a
very simple and transparent description of the effect. We then remove this
approximation and demonstrate that certain RNA dark states do remain
dark in the exact analysis. The decay rates of the other RNA dark states
are estimated, and the analytical results compared to numerical calculations.
A central result is that even though these states are only approximately dark,
they still have extremely long lifetimes. This should render the existence
of the quasi-dark states amenable to experimental observation, at least in
principle. Finally, further remarks and conclusions are given in Sec.~V.

\section{Model}

Our objective consists in studying the center-of-mass motion of two atoms
confined by a trapping potential and interacting with the electromagnetic
field inside a high-$Q$ cavity. In order to work out most clearly
some of the basic physical effects observable in this system we investigate in
the following an idealized model problem. Questions of experimental
realizability will be discussed in Sec.\ V.

We consider the one-dimensional motion of two two-level atoms of mass $M$
trapped inside an infinite square-well potential $V(x)$ with boundaries
at $x=0$ and $x=L$. The upper and lower internal atomic states $|e\rangle$ and
$|g\rangle$ are separated in energy by an amount of $\hbar\omega_0$. The atoms
which are treated as distinguishable are also placed inside a short and
closed electromagnetic cavity that is aligned with the atomic trap along the
$x$ axis. We assume the cavity characteristics to be such that the atomic
interaction with the cavity field can be described as a coupling to a single
mode. In particular, spontaneous photon emission into directions other than the
$x$ axis is disregarded. On the other hand, the damping of the relevant
cavity mode due to its coupling to the electromagnetic vacuum outside
the resonator is taken into account. Based on this description, the
Hamiltonian of the system is
\begin{equation}
H=H_a+H_c+H_r+H_{ca}+H_{cr}
\label{h}
\end{equation}
where $H_a$, $H_c$ and $H_r$ are the free Hamiltonians of the atoms,
the cavity mode and the vacuum modes, respectively. They are given by
\begin{equation}
H_a=\sum_{j=1}^2\left (\frac{\hat{p}_j^2}{2M} + V({\hat x}_j) +
\hbar\omega_0\sigma_j^ {\dagger}\sigma_j \right ),
\label{h1}
\end{equation}
\begin{equation}
H_c=\hbar\omega_c a_c^{\dagger} a_c,\quad H_r=\sum_{\mu}\hbar\omega_{\mu}
a_{\mu}^{\dagger} a_{\mu}.
\label{h2}
\end{equation}
Here, $\hat{p}_j$ is the center-of-mass momentum and ${\hat x}_j$ the 
position of the $j$th atom along the
$x$-axis. The atomic pseudo-spin operators $\sigma_j$ are defined by
$\sigma_j=|g,j\rangle\langle e,j|$.
The annihilation operators for the cavity mode and the vacuum modes are
denoted $a_c$ and $a_\mu$, respectively, the mode frequencies are
$\omega_c$ and $\omega_{\mu}$. The interaction of the cavity mode with the
atoms and with the vacuum modes are described by the terms $H_{ca}$ and
$H_{cr}$. In the dipole and the rotating-wave approximation, they read
\begin{equation}\label{hca}
H_{ca}=\sum_{j=1}^2\hbar g\cos(kx_j+\phi)(\sigma_j^\dagger a_c+
\sigma_ja_c^\dagger),
\label{h3}
\end{equation}
\begin{equation}
H_{cr}=\sum_\mu\hbar(g_\mu^\ast a_c^\dagger a_\mu+g_\mu a_ca_\mu^\dagger),
\label{h4}
\end{equation}
where $g=(\hbar\omega_c/2\varepsilon_0 L_c)^{1/2}$ denotes the
atom-cavity coupling constant with $L_c$ the cavity length. For a planar
cavity the mode profile is cosine-shaped with wavevector $k$. The phase
angle $\phi$ characterizes the relative positioning between 
cavity mode and atomic trap. The coupling constant between the cavity
mode and the $\mu$th vacuum mode is denoted $g_{\mu}$.

In discussing the atomic time evolution we will mostly be concerned with
situations in which the center-of-mass wave function is spread out over
a region of extension $\Delta x$ large in comparison to the cavity mode
wavelength $2\pi/k$ but small in comparison to the trap length $L$. For
small enough times the existence of the trap walls may thus be neglected.
Furthermore, it is assumed that the initial wave function can be
ascribed a well-defined momentum $(p_{01},p_{02})$ and that the
effects of the (small) momentum spread around this initial value may be
disregarded. From the form (\ref{h3}) of the atom-field coupling it follows
that a single-atom state with momentum $p$ is only coupled to states
with momenta $p\pm \hbar k$. In view of our initial condition
we thus introduce the notation $|(i_1,m_1),(i_2,m_2),n_c,\{n_{\mu}\}\rangle$
that denotes a state where atom $j$ has internal state $i_j$ and
momentum $q_{0j}+ m_j\hbar k $ with integer $m_j$. Thereby
, $q_{0j}=\mbox{mod}(p_{0j},\hbar
k)$, i.e., $0\leq q_{0j} < \hbar k$. The number of photons in the cavity
and the vacuum mode ``$\mu$''  are denoted $n_c$ and $n_{\mu}$, respectively. 

In case only one excitation is present in the system and within the realm of
validity of the above approximations, the general expression for the system
state vector is thus given by
\begin{eqnarray}
|\Psi(t)\rangle&=&\sum_{m,n}\large\{C_{1,m,n}(t)|(e,m),(g,n),0,\{0_\mu\}
\rangle\nonumber\\ && +C_{2,m,n}(t)|(g,m),(e,n),0,\{0_\mu\}\rangle\nonumber\\
&&+ C_{3,m,n}(t)|(g,m),(g,n),1,\{0_\mu\}\rangle\nonumber\\
&&+\sum_\mu C_{4,m,n,\mu}(t)|(g,m),(g,n),0,\{1_\mu\}\rangle\large\}.
\label{p1}
\end{eqnarray}
We now proceed to eliminate the reservoir degrees of freedom in the
system equations of motion with the help of the Born-Markov
approximation. This introduces an exponential decay rate $\kappa/2=
\pi|g_{\mu}|^2$ and a frequency shift $\Delta_c$ in the dynamics of the
amplitudes $C_{3,m,n}$. For the following, we incorporate this shift into
the detuning $\Delta$ between the atomic resonance and the cavity frequency
and work in the interaction picture with respect to
$\omega_0$. The effective Hamiltonian time evolution of the system
before the photon escapes the cavity is then determined by
\begin{eqnarray} \label{C11}
i\dot{C}_{1,m,n}&=&\omega_{m,n}C_{1,m,n}+\frac{g}{2}(C_{3,m+1,n}
+ C_{3,m-1,n}),\\
i\dot{C}_{2,m,n}&=&\omega_{m,n}C_{2,m,n}+\frac{g}{2}(C_{3,m,n+1}
+ C_{3,m,n-1}), \label{C0} \\
\label{C22}
i\dot{C}_{3,m,n}&=&(\omega_{m,n}+\Delta-i\kappa/2) C_{3,m,n}
+\frac{g}{2}(C_{1,m+1,n}\\ &&+ C_{1,m-1,n}+
C_{2,m,n+1}+ C_{2,m,n-1})\nonumber
\end{eqnarray}
with
\begin{equation}
\omega_{m,n}=[(q_{01}+m\hbar k)^2+(q_{02}+n\hbar k)^2]/(2\hbar M)
\end{equation}
describing the influence of kinetic energy. From Eqs.\ (\ref{C11})-(\ref{C22}) 
one notices a further selection rule. For example, the set of
coefficients $C_{1,m,n}$ with $m,n$ both even, are only coupled among
each other and to $C_{2,m',n'}$, $m',n'$ odd, and $C_{3,m'',n''}$, $m''$ odd,
$n''$ even. Note also that Eqs.\ (\ref{C11})-(\ref{C22}) can be written
independently of the phase angle $\phi$. In the following we set $\Delta=0$
for convenience.

Another interesting situation arises if one takes the existence of the
atomic trap boundaries fully into account. In this case it is convenient
to expand the center-of-mass wave functions in terms of the eigenfunctions of
the atomic Hamiltonian (\ref{h1}), i.e., $2\sin(\pi q x_1/L) \sin(\pi r x_2/L)/L$,
$q,r\geq 1$, which can be thought of as specific superpositions of momentum 
states with opposite wave vectors. In general, the coupling term $H_{ac}$ 
introduces transitions from a single-particle eigenstate $\psi_q^{g(e)}=
\sqrt{2/L}\sin(\pi q x_1/L)|g(e) \rangle$ to an infinite number of other
states $\psi_{q'}^{e(g)}$. Simple selection rules follow if one has
$k=N\pi/L$ with $N$ a positive integer and $\phi=0$. Under these conditions
one obtains couplings only between the single-atom wave functions
\begin{equation}\label{coup}
\dots \leftrightarrow \psi_{2N-q}^{g/e} \leftrightarrow \psi_{N-q}^{e/g}
\leftrightarrow \psi_q^{g/e} \leftrightarrow \psi_{q+N}^{e/g} \leftrightarrow
\psi_{q+2N}^{g/e} \leftrightarrow \dots\,
\end{equation}
with $1\leq q<N$. The coupling coefficients are all equal besides the one between
$\psi_{N-q}$ and $\psi_q$ which is of the same magnitude, but of opposite
sign. After suitable identifications the equations of motion for the
probability amplitudes of the two-atom system can thus be cast into a form
identical to Eqs.\ (\ref{C11})-(\ref{C22}) apart from this sign peculiarity.
An important special case in the coupling scheme of expression
(\ref{coup}) arises if $q=N$. Under these circumstances the sequence
terminates at $\psi_q$, the part to the left of it being obsolete. This special
case is of particular importance in the discussion of exact dark states
beyond the RNA.

\section{Numerical results}

In order to set the stage for the two-atom problem, let us first take a 
brief look at its one-atom counterpart. With the help of the procedure used 
to derive Eqs.\ (\ref{C11})-(\ref{C22}) we can obtain a similar set of 
equations for the one-atom system,
\begin{equation} \label{CC1}
i\dot{C}_{1,m}=\omega_m C_{1,m}+\frac{g}{2}(C_{2,m+1}+C_{2,m-1}),
\end{equation}
\begin{equation}
i\dot{C}_{2,m}=(\omega_m+\Delta-i\kappa/2)C_{2,m}
+\frac{g}{2}(C_{1,m+1}+C_{1,m-1})
\label{C1}
\end{equation}
where the notations used here are defined in parallel to those for the two-atom
case. In particular, we now have $\omega_m=(q_0+m\hbar k)^2/(2\hbar M)$. The
excited and ground state amplitudes are denoted $C_1$ and $C_2$, respectively.
Equations (\ref{CC1}) and (\ref{C1}) are very similar in structure to those
used in the discussion of near-resonant scattering of two-level atoms from a
standing-wave laser field \cite{Kaz}. Physically, they describe the
atomic momentum spread during the interaction with the cavity mode. If we
imagine the standing wave mode as being composed of two counterpropagating
running waves we see that during an emission-absorption cycle the atomic
momentum can change by an amount of 0 or $2\hbar k$. The change depends
on whether the photon is emitted into and absorbed from the same running
wave mode or not. Successive cycles thus lead to an atomic momentum
spread, i.e.\ heating.

This is illustrated in Fig.\ \ref{fig1}, which shows momentum distributions
$P_m(\tau)=|C_{1,m}(\tau)|^2+ |C_{2,m}(\tau)|^2$ derived from
Eqs.\ (\ref{CC1}) and (\ref{C1}) as a function of the discrete
momentum index $m$ and the dimensionless time $\tau=\omega_{rec}t$, with
$\omega_{rec}=\hbar k^2/(2M)$ being the recoil frequency. These distributions
illustrate the effective Hamiltonian time evolution of the atom
before the photon escapes the cavity, governed by the nonhermitian 
Hamiltonian
\begin{equation}
H_{eff} = H_a + H_c + H_{ca} - i\hbar \frac{\kappa}{2} a_c^\dagger a_c,
\end{equation}
$H_a$ and $H_{ca}$ referring now to a single two-level atom. The initial
conditions
for the wave function were chosen as $C_{1,m}=\delta_{m,0}$, $C_{2,m}=0$,
and $q_0=0$. Figures \ref{fig1}(a),(b) display the case of a lossless
cavity ($\kappa=0$) and a dimensionless atom-cavity coupling constant
$\Omega=g/2\omega_{rec}=50$. In Fig.\ \ref{fig1}(a), the
influence of the kinetic energy term $\hat{p}^2/2M$ is neglected
(the Raman-Nath approximation) and the momentum spread grows linearly in time
at a rate proportional to $\Omega\tau$. This should be compared to Fig.
\ref{fig1}(b), which is for the full model including the kinetic energy
terms. This illustrates the well-known fact that the RNA is
only valid for short enough times. Due to the increasing mismatch between
the photon energy and the atomic energy increment accompanying a photon
absorption, the width of the momentum distribution eventually stops growing
and begins to oscillate. The effects of cavity damping are illustrated in
Fig.\ \ref{fig1}(c) and (d), which again compare the momentum distributions
in the RNA and the full model, but for a moderate cavity damping rate
$\kappa'=\kappa/\omega_{rec}=50$, i.e., $\kappa'
/\Omega=0.4$. In this case the excited state
population is damped on a time scale approximately given by
$4/\kappa'$.\footnote{It should be noted that for large cavity damping
$\kappa'\gg\Omega/2$ the decay rate of the excited state population goes to
zero. This stabilization effect, however, is different in nature from
the two-atom dark states discussed below.}

We now turn to the two-atom situation, with the goal of determining how
the previous results are modified when we insert a second atom into the cavity.
The dramatic changes brought about under these circumstances are illustrated in
Figs.\ \ref{fig2}(a)-(d), which show results of the numerical integration of
Eqs.\ (\ref{C11})-(\ref{C22}). They depict the momentum distribution of the
first atom before the photon escape,
$P^{(1)}_m(\tau)= \sum_{i=1,2,3;n} |C_{i,m,n}(\tau)|^2$, as a function
of $m$ and $\tau$ in both the RNA and the full model,
and in the absence or presence of cavity losses. The initial conditions
were chosen such that both atoms are at rest but
atom 1 is in the excited state, atom 2 is in the ground state and no
photon is present in either the cavity or the vacuum modes, i.e.,
\begin{equation}
C_{i,m,n}(t=0)=\delta_{i,1}\delta_{m,0}\delta_{n,0}
\end{equation}
and $p_{01}=p_{02}=0$. The atom-cavity coupling is again set to $\Omega=50$. 
As a consequence of the selection rules mentioned in Sec.\ II one has for
these initial conditions
$$P^{(1)}_m=\sum_n |C_{1,m,n}|^2$$
for $m$ even and
$$P^{(1)}_m=\sum_n |C_{2,m,n}|^2+|C_{3,m,n}|^2$$
for $m$ odd. Figures \ref{fig2}(a) and (b) display the case of the lossless
cavity. One can recognize two main qualitative differences from the
corresponding Figs.\ \ref{fig1}(a) and (b). First, the momentum distribution 
no longer spreads significantly: rather, it remains
concentrated in the central mode (i.e.\ $m=0$) and a small number of side
modes. The other modes remain almost unpopulated. Second, the
comparison between the RNA and the full model results shows that the influence
of the kinetic energy terms now is much smaller than in the one-atom case.
Contrary to Figs.\ \ref{fig1}(a) and (b), for the time considered they
only lead to some quantitative modifications but not to a qualitative
change. This property is of course due to the concentration of the
momentum distribution around $m=0$. It also indicates that the RNA is a
valuable tool in the interpretation of the two-atom behavior.

The study of the momentum distribution in the presence of cavity losses
[Figs.\ \ref{fig2}(c) and (d), again with $\kappa'=20$] also yields a
surprising result. One finds again that only a small number of modes are
significantly populated. But in addition, and in contrast to the one-atom
case, after an initial transient evolution the total atomic population decays
only very slowly, i.e., {\em the photon escape from the cavity is strongly
inhibited by the presence of a second atom.} In fact, the time evolution of 
the distribution still bears a strong similarity to the lossless case.
Furthermore, the RNA yields a good approximation to the full model also
in the presence of losses. A further increase of the cavity damping rate
only leads to minor changes in the behavior of the momentum distribution. 

A closer look at the long-time behavior is provided in Figs.\ \ref{fig3}.
There, the total probability $P=\sum_m P_m^{(1)}$ of finding the
excitation in the cavity (curve 1) is shown for the RNA (a) and the full
model (b). The parameter values are chosen as in Figs.\ \ref{fig2}(c),(d).
After a rapid initial transient the probability $P$ reaches a constant value
in the RNA, whereas it still decays slowly in the full model. The curves 2
and 3 show the time evolution of $|C_{1,0,0}|^2+
|C_{1,0,\pm 2}|^2+|C_{1,\pm 2,0}|^2+|C_{2,\pm 1,\pm 1}|^2$ (i.e., the
central and the most highly populated side modes)and of $|C_{1,0,0}|^2$
alone, respectively. These curves
again demonstrate that the spread in momentum is strongly suppressed.

\section{Two-atom dark states}

The results of Figs.\ \ref{fig2} and \ref{fig3} indicate that the
atomic time evolution is characterized by the appearance of dark states
which have the initial excitation stored in the atoms and which are
almost immune to cavity damping. In this section a
detailed analysis of these dark states is given. Before turning to the
full problem we first work in the RNA, which was shown to provide a
useful approximate description.

\subsection{Two-atom dark states in the Raman-Nath approximation}

In order to investigate the dark states it is convenient to work also
in the position-space representation. The equations of motion for the
position-dependent probability amplitudes $C_i(x_1,x_2,t)$ read
\begin{eqnarray}
i\dot{C}_1&=&-\frac{\hbar}{2M}\left(\frac{\partial^2}{\partial x_1^2}+
\frac{\partial^2}{\partial x_2^2}\right)C_1+g\cos(kx_1)C_3 \label{x1},\\
i\dot{C}_2&=&-\frac{\hbar}{2M}\left(\frac{\partial^2}{\partial x_1^2}+
\frac{\partial^2}{\partial x_2^2}\right)C_2+g\cos(kx_2)C_3 \label{x2},\\
i\dot{C}_3&=&\left[-\frac{\hbar}{2M}\left(\frac{\partial^2}{\partial x_1^2}+
\frac{\partial^2}{\partial x_2^2}\right)+\Delta-i\kappa/2\right]C_3
\nonumber \\
&& +g\left[\cos(kx_1)C_1+\cos(kx_2)C_2\right]. \label{x3}
\end{eqnarray}
In the first special case discussed in Sec.\ II (i.e., the atomic wave
packet well localized inside the trap) these equations have to be solved
in the domain $0\leq x_1,x_2 \leq 2\pi/k$ and the solution must be of the
form
\begin{equation}
C_i=\exp(ip_{01}x_1+ip_{02}x_2)\tilde{C}_i
\end{equation}
with $\tilde{C}_i$ fulfilling periodic boundary conditions. In the second
case (trap conditions taken fully into account) one has to consider solutions
with vanishing Dirichlet boundary conditions in the domain
$0\leq x_1,x_2 \leq L$.

In the RNA, i.e., after discarding the spatial derivatives, Eqs.\
(\ref{x1})-(\ref{x3}) decouple spatially and can be solved immediately. At a
given point $(x_1,x_2)$ they form a homogeneous linear $3\times 3$-system of
ordinary differential equations the eigenvalues of which are given by
\begin{eqnarray}
\lambda_1&=&0,\\
\lambda_{2,3}&=&-\kappa/4-i\Delta/2 \label{lam23}\\ && \pm
\sqrt{\left(\frac{\kappa}{4}+i\frac{\Delta}{2}\right)^2-g^2
[\cos^2(kx_1)+\cos^2(kx_2)]}.\nonumber
\end{eqnarray}
The existence of the eigenvalue $\lambda_1$ whose real part vanishes
independently of the values of $x_1$, $x_2$, and $\kappa$ ensures that an
excitation initially present in the system has a finite probability of
remaining in it in the limit $t\to\infty$. In particular, if the atomic
wave function is given at time $t=0$ by
\begin{eqnarray}
|\psi(x_1, x_2,0) \rangle &=& A_1(x_1,x_2) |e,g,0,\{0_\mu \} \rangle 
\nonumber \\ &&+
A_2(x_1,x_2)|g,e,0,\{0_\mu\}\rangle \nonumber \\
&&+ A_3(x_1,x_2)|g,g,1,\{0_\mu\}\rangle
\end{eqnarray}
then the asymptotic state reached by the ``atoms + cavity mode'' system
is characterized by the probability amplitudes (arranged as a column vector
in a self-evident way)
\begin{eqnarray}\label{assta}
&&[\cos^2(kx_1)+\cos^2(kx_2)]^{-1}\\ && \times \left(\begin{array}{c}
A_1\cos^2(kx_2)-A_2\cos(kx_1)\cos(kx_2)\\
-A_1\cos(kx_1)\cos(kx_2)+A_2\cos^2(kx_2)\\
0 \end{array} \right)\nonumber.
\end{eqnarray}
Note that this state is {\em not} normalized, a result of the
fact that some of the initial excitation has irreversibly escaped from the
cavity into the reservoir.

Expression (\ref{assta}) shows that the asymptotic state
does not have a contribution from the initial amplitude $A_3$,
furthermore, the final amplitude in the third channel where the photon
is present in the cavity vanishes. On the other hand, if a state has
nonvanishing contributions $A_1$ or $A_2$ it will always evolve into a
dark state unless $A_1\cos(kx_2)=A_2\cos(kx_1)$. The time scale to reach
the dark state is determined by the eigenvalues $\lambda_2$ and $\lambda_3$.

From Eqs.\ (\ref{x1})-(\ref{x3}) or Eq.\ (\ref{assta}) it follows that a
given state is a dark state if and only if it is of the form
\begin{equation}\label{dsta}
A(x_1,x_2)\left(\begin{array}{c}
\cos(kx_2) \\ -\cos(kx_1) \\ 0 \end{array} \right)
\end{equation}
and, in addition, it fulfills the appropriate boundary conditions. The state
(\ref{dsta}) can be viewed as a generalization of the dark state in the
Dicke theory of sub- and superradiance \cite{Dic54}.

In the following discussion we concentrate on the case of localized atoms
in the sense of Sec.\ II.  If one substitutes for the function $A$ of
expression (\ref{dsta}) the set of plane waves
$\exp(iq_{01}x_1+iq_{02}x_2) \exp[imkx_1+i(n+1)kx_2]$, one obtains a family
of dark states $\{ |d_{mn} \rangle\}$ which have a simple structure in
momentum space, i.e.,
\begin{eqnarray}
|d_{mn}\rangle&&=\textstyle{\frac 1 2}\large(|(e,m),(g,n)\rangle + |(e,m),
(g,n+2)\rangle \\
&-&|(g,m+1),(e,n+1)\rangle -|(g,m-1),(e,n+1)\rangle\large),\nonumber
\end{eqnarray}
where we have omitted the occupation numbers of the photon modes in the
notation of the ket vectors for simplicity. The dark states
$|d_{mn}\rangle$ are truly entangled states. Since all permissible
functions $A$ can be expanded onto the indicated set of plane waves
the family $\{ |d_{mn}\rangle\}$ forms a basis of the ``dark'' subspace of
the total Hilbert space. However, this is not an orthogonal basis as a
given $|d_{mn}\rangle$ has a nonvanishing scalar product with four
other $|d_{m'n'}\rangle$.

Of particular interest in our context is the question of how to characterize
the asymptotic state $|D_{mn}^{e/g,g/e}\rangle$ associated with a given initial
state $|(e/g,m),(g/e,n)\rangle$. Its coordinate representation can be
inferred immediately from Eq.\ (\ref{assta}), but further
insight into the nature of the state can be obtained from its
momentum distribution. Equations (\ref{C11})-(\ref{C22}) show that it is
sufficient to study this question for the state $|D_{00}^{eg}
\rangle$, since the distributions for the other states can be obtained
by a suitable shift of indices. In coordinate space the state $|D_{00}^{eg}
\rangle$ is represented by
\begin{eqnarray}
&&[\cos^2(kx_1)+\cos^2(kx_2)]^{-1}\\
&&\times(\cos^2(kx_2),-\cos(kx_1)\cos(kx_2),0)^{T}. \nonumber
\end{eqnarray}
Its momentum-space amplitudes
$$c_{1/2,m,n}= \langle (e/g,m),(g/e,n)| D_{00}^{eg}\rangle$$
are determined by
\begin{eqnarray}\label{cint}
c_{i,m,n}=\frac{k}{2\pi}\int \int_0^{2\pi /k}\!\! dx_1dx_2\, e^{-i(mkx_1+nkx_2)}
\nonumber \\
\times\frac{f_i(x_1,x_2)}{\cos^2(kx_1)+\cos^2(kx_2)}
\end{eqnarray}
with $f_1=\cos^2(kx_2)$ and $f_2=-\cos(kx_1)\cos(kx_2)$. As discussed in
Sec.\ II, $c_{1(2),m,n}\neq 0$ only for $m,n$ both even (odd).
Evaluating the integrals (\ref{cint}) one finds that the amplitudes
$c_{1,2m,0}$, $m\geq 0$, are given by
\begin{equation}\label{rec}
c_{1,2m,0}=\delta_{m,0}+\frac{i}{2\pi}(I_m +I_{m-1})
\end{equation}
where the numbers $I_m$ satisfy the recurrence relation
\begin{equation}
I_m=\frac 1 m[(-1)^{m-1}4i -(6m-3)I_{m-1} -(m-1)I_{m-2}]
\end{equation}
and $I_{0}=I_{-1}=i\pi/2$. Further relations between the  amplitudes
$c_{i,m,n}$ are given by
\begin{eqnarray}
&&c_{1,m,n}+c_{1,m+2,n}+c_{2,m+1,n+1}+c_{2,m+1,n-1}=0, \label{rel1}\\
&&c_{1,m,n}+c_{1,m,n+2}-c_{2,m+1,n+1}-c_{2,m-1,n+1} \nonumber\\
&&\quad =\delta_{m,0}(\delta_{n,0} +\delta_{n,-2}), \label{rel2}\\
&&c_{i,m,n}=c_{i,\pm m,\pm n} \label{rel3}
\end{eqnarray}
with $m,n$ both even in Eqs.\ (\ref{rel1}) and (\ref{rel2}). Equation
(\ref{rel1}) is a direct consequence of Eq.\ (\ref{C22}) whereas Eq.
(\ref{rel2}) follows from the relation $$|d_{00}\rangle=(|D_{00}^{eg}
\rangle+|D_{02}^{eg}\rangle-|D_{11}^{ge}\rangle-|D_{-1,1}^{ge} \rangle)/2.
$$ With the help of Eqs.\ (\ref{rec})-(\ref{rel3}) all amplitudes $c_{i,m,n}$
can be calculated iteratively. In this way, one obtains for example
\begin{eqnarray*}
c_{1,0,0}&=&1/2,\\
c_{2,\pm 1,\pm 1}&=&1/\pi-1/2\approx -0.1817, \\
c_{1,\pm 2,0}&=&-c_{1,0,\pm 2}=1/2-2/\pi\approx -0.1366.
\end{eqnarray*}

An interesting way to determine the scalar products $\langle D_{m,n}^{\sigma}
|D_{00}^{eg}\rangle $ with $\sigma=eg$ or $ge$ proceeds as follows
[the method can also be used to derive
Eq.\ (\ref{rel2})]. The asymptotic state
$|D_{00}^{eg}\rangle$ into which $|(e,0),(g,0)\rangle$ evolves is
uniquely determined. Any state in the ``dark subspace" orthogonal to
$|D_{00}^{eg}\rangle$ must have vanishing overlap with
$|(e,0),(g,0)\rangle$. If we denote by $|\bar{D}_{00}^{eg}\rangle$ the
state $|D_{00}^{eg}\rangle$ after normalization --- remember that the
dark state into which a given initial state evolves is not normalized ---
we must have that
$$
|D_{00}^{eg}\rangle=|\bar{D}_{00}^{eg}\rangle
\langle \bar{D}_{00}^{eg}|(e,0),(g,0)\rangle.
$$
Comparing coefficients one obtains that
\begin{equation}\label{sc1}
\langle D_{00}^{eg} |D_{00}^{eg}\rangle = 0.5,
\end{equation}
i.e., the system has a 50$\%$ probability to be trapped in that dark state.
Using the Gram-Schmidt orthogonalization scheme to construct from
$|D_{m,n}^{\sigma}\rangle$ a state orthogonal to $|\bar{D}_{00}^{eg}
\rangle$ leads to the conclusion that
\begin{equation} \label{sc2}
\langle D_{m,n}^{\sigma}|D_{00}^{eg}\rangle =c_{k,m,n}
\end{equation}
with $k=1(2)$ if $\sigma=eg(ge)$,
i.e., the asymptotic dark states are non-orthogonal, in general.
Equations (\ref{sc1}) and (\ref{sc2}) can be verified by evaluating
the scalar product in position space.

From Eqs.\ (\ref{rec})-(\ref{sc1}) it can be inferred that 50$\%$ of the
population of the dark state is trapped in the state $|(e,0),(g,0)\rangle$,
while the states $|(i,m),(j,n)\rangle$ with $|m|+|n|\leq 2$ $(4)$ hold
91.3$\%$ ($96.3\%$) of the population. This observation explains the
localization of the momentum distributions in Figs.\ \ref{fig2} and
\ref{fig3}.

\subsection{Exact and approximate dark states in the full model}

Turning to the full model described by Eqs.\ (\ref{C11})-(\ref{C22}) or
(\ref{x1})-(\ref{x3}), i.e., taking the kinetic energy terms into
account, it becomes apparent that, in general, the states $|d_{mn}\rangle$ and
$|D_{mn}\rangle$ are no longer exactly dark. By `exactly dark' we mean being
an eigenstate of the full Hamiltonian with a purely real eigenvalue. It is
therefore natural to ask whether the full model sustains exact dark
states at all. Interestingly, a complete answer to this question can be
given for both cases discussed in Sec.\ II, i.e., for atoms localized
well inside the trap and for atoms experiencing the trap boundaries.
In the first situation there are precisely two exact dark states, which
are given by
\begin{equation}
|D_1\rangle=|d_{0,-1}\rangle=(\cos(kx_2),-\cos(kx_1),0)^T
\end{equation}
and
\begin{eqnarray}\label{da2}
|D_2\rangle&=&\sin(kx_1)\sin(kx_2)(\cos(kx_2),-\cos(kx_1),0)^T\nonumber \\
&=&|d_{-1,0}\rangle-|d_{1,0}\rangle+|d_{1,-2}\rangle-|d_{-1,-2}\rangle.
\end{eqnarray}
Dark states thus appear only if the atomic momenta involved  are integer
multiples of $\hbar k$, i.e., if $q_{01}=q_{02}=0$. For the second case, in
which the atomic wave functions extend over the whole length of the trap,
it can be shown that exact dark states can only exist if in the cavity mode 
function of Eq.\ (\ref{hca}) $k=\pi N/L$ with integer $N\geq 1$ and $\phi=0$. Under
these conditions there is precisely one such state which, in the
coordinate representation, is given by the first line of Eq.\ (\ref{da2}).

For a proof of uniqueness of these dark states one can start from the
observation that also in the full model exact dark states have to be of
the form (\ref{dsta}). Additionally, they now also must be
eigenfunctions of $(\hat{p}^2_1 +\hat{p}^2_2)/2M$ under the
appropriate boundary conditions. One then expands both $A(x_1,x_2)$ and
$A(x_1,x_2)\cos(kx_{1/2})$ onto a suitable set of eigenfunctions. The
fact that in the expansion of $A(x_1,x_2)\cos(kx_{1/2})$ there should
only appear terms of the same energy imposes severe restrictions on the
possible forms for the expansion of $A(x_1,x_2)$. These requirements can
only be met in the cases indicated. For the situation in which the atoms
extend over the whole trap the breakoff of the coupling scheme
(\ref{coup}) if $q=N$ (as outlined at the end of Sec.\ II) turns out to
be crucial for the existence of the dark state.

These considerations imply that most dark states found in the RNA
become unstable in the full model since they are orthogonal to the exact
dark states, in general. The numerical results of Sec.\ III suggest,
however, that the corresponding lifetimes are still very long so that
these states may be regarded as ``quasi-dark." The examples shown referred to
cases in which $\Omega,\kappa' \gg 1$ which is the relevant situation in
practice as discussed in Sec.\ V. Under these conditions one may treat
the kinetic energy term $(\hat{p}_1^2+\hat{p}_2^2)/2M$ as a small
perturbation to the RNA Hamiltonian. Applying standard perturbation
theory one obtains an imaginary correction to the RNA dark state
eigenenergies only in second order, which already indicates that these
states will be long-lived. A crude estimate of the second-order
imaginary part shows that the state
$|D(d)_{mn}\rangle$ acquires a finite decay rate that is of the order of
\begin{equation}
\Gamma_{mn}  \simeq \omega_{rec}(\tilde m^2+\tilde n^2)^2\kappa'/\Omega^2 .
\label{estimate}
\end{equation}
Thereby, $\tilde m$ and $\tilde n$ have to be understood as typical
values of $m$ and $n$ appearing in the expansion into center-of-mass
momentum states.  The estimate (\ref{estimate}) assumes that
$\kappa'$ is not too large in comparison to $\Omega$ so that the square
root in expression (\ref{lam23}) is essentially imaginary.

Hence, consistently with the numerical calculations, we find that the
lifetime of the
`quasi-dark states' is long compared to $\omega_{rec}^{-1}$ under the
condition $\Omega,\kappa' \gg 1$. Furthermore, our estimate implies that
the decay rate increase rapidly for increasing $m,n$. This is as can be
expected,  since under these circumstances the dephasing
between the different momentum eigenstates becomes faster.
The dependence on $\kappa'$ and $\Omega$ suggest that the coupling to
the decay channel becomes more efficient when $\kappa'$ is increased and
$\Omega$ decreased. Figure \ref{fig4} shows the decay of the dark states
$|d_{mn}\rangle$ for various values of $(m,n)$, $\kappa'$, and $\Omega$.
Their evolution qualitatively confirms the dependence (\ref{estimate}) of
$\Gamma_{mn}$ on these parameters. Thereby, curve (a) should be compared
to curves (b), (c), and (d) as in each one of these one relevant parameter
is changed in comparison to (a).

\section{Summary and conclusions}

In this paper we have investigated the dynamics of two two-level atoms coupled
to a single damped mode of an electromagnetic resonator, including the 
effects of photon recoil. We concentrated on the situation where one quantum
of excitation is initially present in the system. A generic feature of the 
atomic evolution is the appearance of dark states. These states, in which 
the excitation is stored in the internal atomic degrees of freedom, are
almost immune to photon decay from the cavity. When in a dark state, the 
two atoms become quantum mechanically entangled and form a new kind
of ``molecule'' bound by the quantum of excitation that they share. The state
of the compound system can conveniently be described in terms of a
superposition of different states of well-defined center-of-mass momentum.
A remarkable characteristic feature of the dark states is their small 
momentum spread, as compared e.g. to the one-atom situation. This property
makes their description in the Raman-Nath approximation quite accurate.
While most dark states become only ``quasi-dark'' when this approximation
is removed, their damping rate remains quite long indeed.

When considering the possible practical realization of these states, an
interesting question concerns the influence of a non-constant atomic 
trapping potential on the time evolution of the dark states. If the trapping 
potentials can be arranged to be equal for ground and excited states,  then 
one can still obtain dark states in the RNA (for the full model it can be
anticipated that exact dark states will not exist any longer, in
general). If, as is normally the case, these potentials differ from each
other, even the RNA will not support dark states. However, as Eqs.\
(\ref{x1})-(\ref{x3}) show, in the vicinity of the line $x_1=x_2$ the decay
will be significantly decelerated so that a remnant of the dark-state
effect might still be visible under such circumstances.

Let us conclude with a brief discussion of the experimental feasibility
to observe such two-atom dark states. Recent cavity QED experiments
in the microwave and optical domain are described e.g., in Refs.\
\cite{Bru96,MieFosOro98,HooChaLyn98}. They typically involve a low density
atomic beam passed through the electromagnetic resonator, a situation that
can be modeled in terms of the localized wave packet description of Sec.\ II.
In these experiments the residual spontaneous atomic decay rate $\gamma$
in the cavity (due to coupling to vacuum modes) is approximately one order
of magnitude smaller than the cavity Rabi frequency $g$ and damping rate
$\kappa$, which are both comparable in magnitude. A single-mode description is
thus adequate and our system (once prepared in the initial state) would have
enough time to coherently evolve into a dark state. Furthermore, the recoil
frequency $\omega_{rec}$ is also very small in comparison to $g$ and
$\kappa$ (typically less than a factor of $10^{-3}$) so the RNA should
provide a very accurate description. In an experimental realization a
main difficulty would certainly consist in efficiently preparing the
initial system state. From this point of view, the optical regime does not
appear as promising as the microwave regime: First, due to the short
free-space spontaneous lifetime of optical transitions the atoms probably
could not be prepared in the excited state before they enter the cavity. Second, 
if they are both simultaneously excited inside the cavity the probability of
coupling to the dark state is relatively low. 

An experiment involving a
microwave cavity might proceed as follows. Diatomic molecules in a
low-intensity beam are dissociated such that the two fragments are of
nonvanishing opposite spin. The atoms can thus be separated in an inhomogeneous 
magnetic field. One atomic beam is subsequently prepared in the Rydberg 
ground state, the other one in the excited state. Using atom optical elements
the two beams are guided such that they intersect each other in the
microwave cavity (at a small angle). As the molecular dissociation creates
atom pairs it should be possible to arrange the setup so that both partners
pass the cavity simultaneously with high probability. The experimental
parameters should be chosen such that a single atom always leaves the cavity in
the ground state. The signature of the formation of a dark state would
consist in detecting an appreciable fraction of atoms leaving the cavity in
the excited state. In order to obtain more information about the nature
of the dark state one could for example additionally observe the spatial 
atomic density distribution.

\acknowledgments
We have benefited from numerous discussions with Dr.\ E.\ V.\ Goldstein
and M.\ G.\ Moore. G.\ J.\ Y.\ gratefully acknowledges support from the
Chinese Scholarship Committee. This work was also supported by the U.S.\
Office of Naval Research under Contract No.\ 14-91-J1205, by the
National Science Foundation under Grant No.\ PHY95-07639, by the
U.S.\ Army Research Office, and by the Joint Services Optics Program.

\begin{figure}
\caption{Time evolution of the single-atom momentum distribution $P_m$
for parameter values $\Delta=0$, $\Omega=50$.
Initially, the atom is at rest in the excited state,
no photons are in the cavity and the vacuum. (a) RNA and $\kappa'=0$,
(b) full model and $\kappa'=0$, (c) RNA but $\kappa'=20$, (d) full model and
$\kappa'=20$.} \label{fig1}
\end{figure}

\begin{figure}
\caption{Time evolution of the momentum distribution $P_m^{(1)}$ for the
first atom in the two-atom problem. Initially, both atoms are at rest and
atom 1 is excited, no photons are in the cavity and the vacuum. Parameter
values and use of RNA for (a)-(d) are the same as in Fig.\ \protect\ref{fig1}.}
\label{fig2}
\end{figure}

\begin{figure}
\caption{Time evolution of the total probability $P(\tau)$ to find the
excitation in the cavity (curve 1), of $|C_{1,0,0}|^2+
|C_{1,0,\pm 2}|^2+|C_{1,\pm 2,0}|^2+|C_{2,\pm 1,\pm 1}|^2$ (curve 2), i.e.,
the central mode and the most highly populated side modes,
and of $|C_{1,0,0}|^2$ alone (3). The parameter values are $\Delta=0$,
$\Omega=50$, $\kappa'=20$. (a) RNA, (b) full model.} \label{fig3}
\end{figure}

\begin{figure}
\caption{Total survival probability $P(\tau)$ for initial states
$|d_{mn}\rangle$ in the full model under various conditions.
Parameter values (a) $m=n=0$, $\kappa'=20$, $\Omega=100$;
(b) $m=n=0$, $\kappa'=100$, $\Omega=50$; 
(c) $m=n=0$, $\kappa'=20$, $\Omega=25$; (d) $m=0$, $n=2$, $\kappa'=20$, 
$\Omega=50$.}  
\label{fig4}
\end{figure}

\begin{references}
\bibitem{Mey92} P.~Meystre, in {\it Progress in Optics 30}, edited by
E.~Wolf (Elsevier, New York, 1992).
\bibitem{Har92} S.~Haroche, in {\it Fundamental Systems in Quantum
Optics}, edited by J.~Dalibard, J.-M.~Raimond, and J.~Zinn-Justin
(North-Holland, Amsterdam, 1992).
\bibitem{Pur46} E.~M.~Purcell, Phys.\ Rev.\ {\bf 69}, 681 (1946).
\bibitem{Kle81} D.~Kleppner, Phys.\ Rev.\ Lett.\ {\bf 47}, 233 (1981).
\bibitem{Bru96} M.~Brune, F.~Schmidt-Kaler, A.~Maali, J.~Dreyer,
E.~Hagley, J.~M.~Raimond, and S.~Haroche, Phys.\ Rev.\ Lett.\ {\bf 76},
1800 (1996).
\bibitem{Eic93} U.~Eichmann, J.~C.~Bergquist, J.~J.~Bollinger,
J.~M.~Gilligan, W.~M.~Itano, D.~J.~Wineland, and M.~G.~Raizen, Phys.\
Rev.\ Lett.\ {\bf 70}, 2359 (1993).
\bibitem{DeV96} R.~G.~DeVoe and R.~G.~Brewer, Phys.\ Rev.\ Lett.\ {\bf
76}, 2049 (1996).
\bibitem{Koc95} P.~Kochan, H.~J.~Carmichael, P.~R.~Morrow, and
M.~G.~Raizen, Phys.\ Rev.\ Lett.\ {\bf 75}, 45 (1995).
\bibitem{Mey97} G.~M.~Meyer and G.~Yeoman, Phys.\ Rev.\ Lett.\ {\bf 79},
2650 (1997).
\bibitem{Rud98} T.~Rudolph and Z.~Ficek, Phys.\ Rev.\ A {\bf 58}, 748
(1998).
\bibitem{Yeo98} G.~Yeoman and G.~M.~Meyer, Phys.\ Rev.\ A {\bf 58}, 2518
(1998).
\bibitem{Gol97} E.~V.~Goldstein and P.~Meystre, Phys.\ Rev.\ A {\bf
56}, 5146 (1997).
\bibitem{Dic54} R.~H.~Dicke, Phys.\ Rev.\ {\bf 93}, 99 (1954).
\bibitem{Kaz} A.~P.~Kazantsev, G.~I.~Surdutovich, and V.~P.~Yakovlev,
{\em Mechanical Action of Light on Atoms} (World Scientific, Singapore, 1990).
\bibitem{MieFosOro98} S.~L.~Mielke, G.~T.~Foster, and L.~A.~Orozco,
Phys.\ Rev.\ Lett.\ {\bf 80}, 3948 (1998).
\bibitem{HooChaLyn98} C.~J.~Hood, M.~S.~Chapman, T.~W.~Lynn, and
H.~J.~Kimble, Phys.\ Rev.\ Lett.\ {\bf 80}, 4157 (1998).
\end{references}
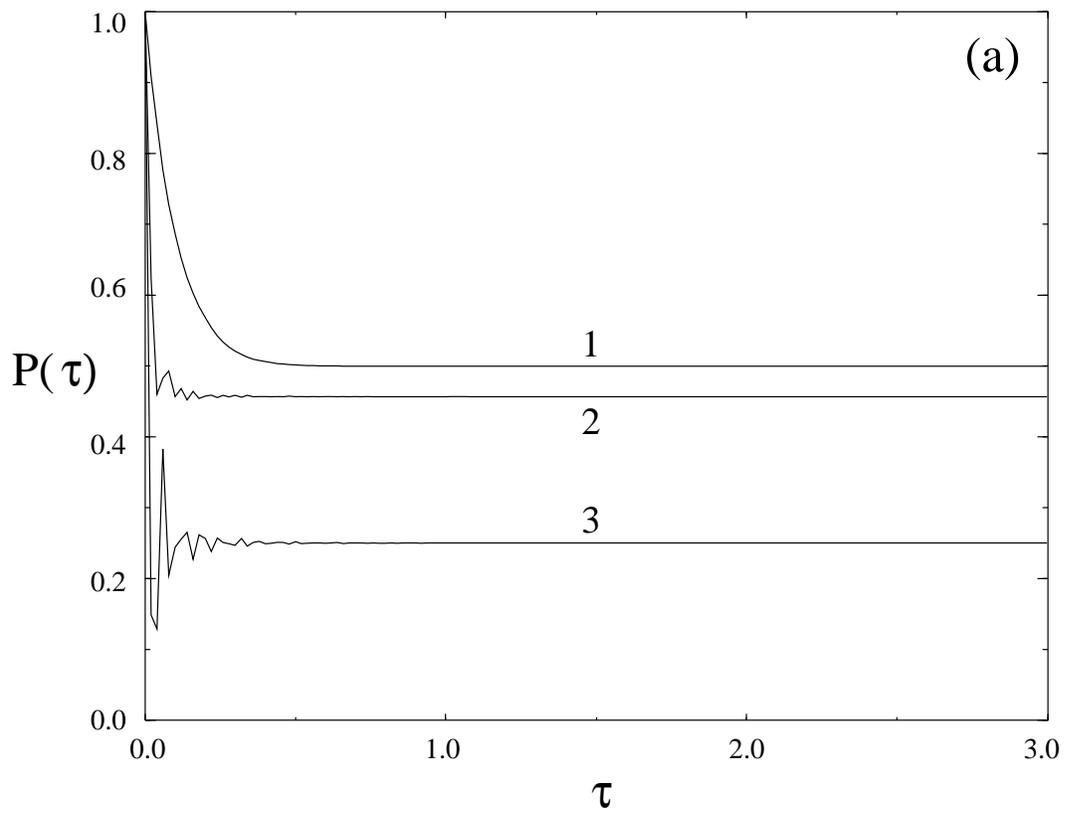
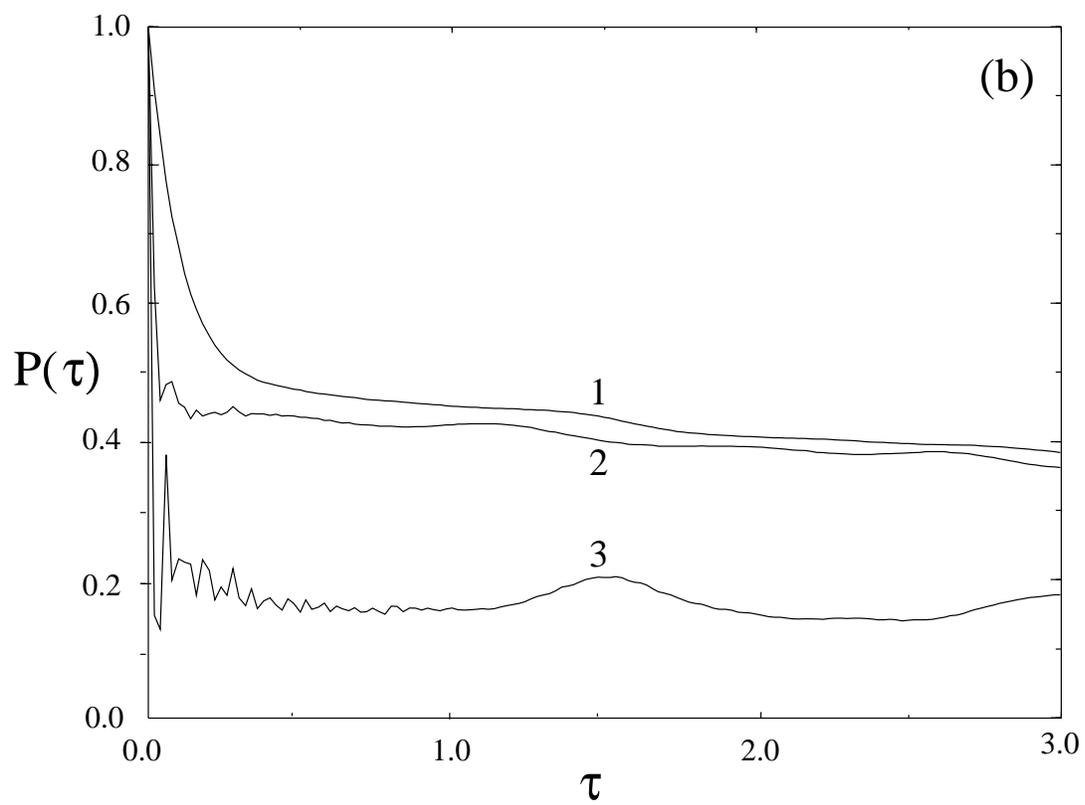
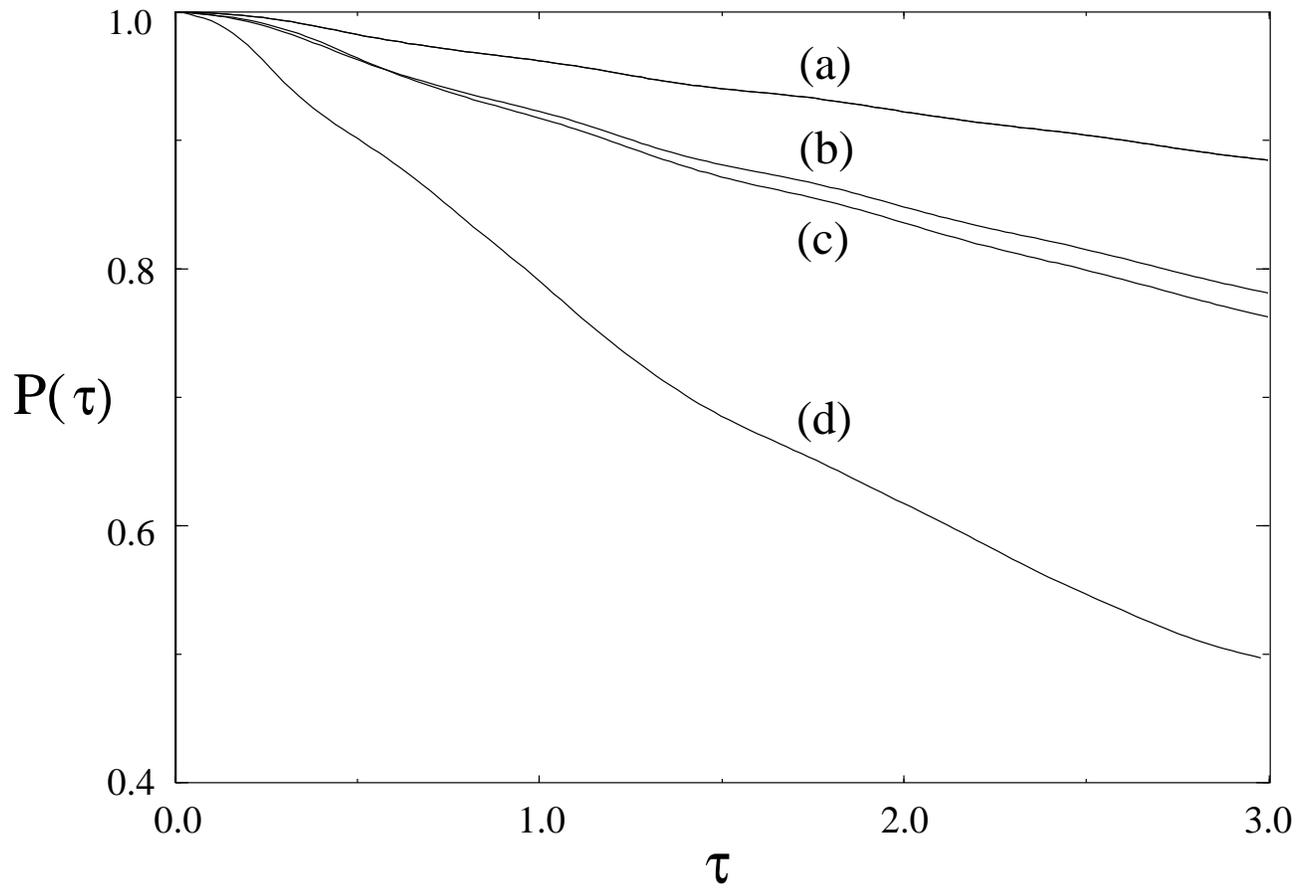
\end{document}